\documentclass[twocolumn,showpacs,prl]{revtex4}
\usepackage{graphicx}
\usepackage{amsfonts}

\begin{document}

\title{Global firing induced by noise or diversity in excitable media}
\author{C.J. Tessone, A. Scir\`e, R. Toral, P. Colet}
\affiliation{Institut Mediterrani d'Estudis Avan\c{c}ats (IMEDEA), CSIC-UIB, Ed.Mateu Orfila, Campus UIB, E-07122 Palma de Mallorca (Spain)}
\begin{abstract} 
We develop a theory for the emergence of global firings in non-identical 
excitable systems subject to noise. Three different dynamical regimes arise: 
sub-threshold motion, where all elements remain confined near the fixed 
point; coherent pulsations, where a macroscopic fraction
fire simultaneously; and incoherent pulsations, where units fire
in a disordered fashion. We also show that the mechanism for global
firing is generic: it arises from degradation of entrainment originated 
either by noise or by diversity.
\end{abstract}
\date{\today}

\maketitle

Excitable behavior appears in a large variety of physical, chemical and
biological systems \cite{excitability,excitability2}. Typically this behavior occurs for
parameter values close to an oscillation bifurcation, and is characterized by a
nonlinear response to perturbations of a stationary state: while small
perturbations induce a smooth return to the fixed point, perturbations exceeding
a given threshold induce a return through a large phase space excursion (firing),
largely independent of the magnitude of the perturbation. Furthermore, after
one firing the system cannot be excited again within a refractory period of time.
In many situations of interest, the firings are induced by random perturbations
or noise~\cite{excitability_noise}.

In coupled excitable systems, global firing (a macroscopic fraction of the units fires
simultaneously) excited by noise has been observed in chemical excitable media
\cite{noisewave}, neuron dynamics \cite{act_neuron} and electronic systems \cite{electronic},
and it has been  described through several theoretical approaches \cite{zaks,park}. This
synchronized firing can be considered as a constructive effect induced by the noise. Other
examples in which noise actually helps to obtain a more ordered behavior are stochastic
resonance \cite{srrmp}, stochastic coherence (or coherence resonance) \cite{PK97}, and
noise-induced phase transitions \cite{bpt94}.

Diversity, the fact that not all units are identical, is an important ingredient in realistic
modeling of coupled systems.   Ensembles of coupled oscillators with diversity have been
paradigmized \cite{kmoto84} and largely studied \cite{strogatz,strogatz2}, with the result
that synchronized behavior can appear once the disorder induced by the diversity is overcome
by the entraining effect of the coupling. It has been shown that in a purely deterministic
excitable system  diversity may induce global collective firing \cite{julyan} if a fraction
of the elements are above the oscillatory bifurcation. So, diversity and noise might be
expected to play a similar role. 

In this work, we develop an analytical understanding for the emergence of global firing in
coupled excitable systems in presence of disorder, either noise or  diversity. We show that
three different dynamical regimes are possible:  sub-threshold motion, where all elements
remain confined near the fixed point;  coherent pulsations, where a macroscopic fraction fire
simultaneously; and  incoherent pulsations, where units fire in a disordered fashion. 
Remarkably, the coherent behavior appears through a genuine phase transition when the noise
intensity, the coupling or the diversity cross a critical value. A second phase transition to
the disordered (incoherent) phase is recovered for large enough noise intensity or diversity,
or small enough coupling.  The mechanism for global firing is the degradation of entrainment
which can be  originated either by noise or diversity.  This is generic and opens a new
scenario  for experimental observations. 

We consider as a prototypical model an 
ensemble of globally coupled {\it active-rotators} $\phi_j(t)$, $j=1,\dots,N$, whose dynamics is given by \cite{themodel}
\begin{equation}\label{eq:phidot}
\dot{\phi_j} = \omega_j -  \sin \phi_j + \frac{C}{N} \sum_{k=1}^N 
\sin\left(\phi_k-\phi_j \right) + \sqrt D \xi_j.
\end{equation}
The natural frequencies $\omega_j$ are distributed according to a probability density function $g(\omega_j)$, 
with mean value $\omega$ and variance $\sigma^2$. Notice that $\omega_j<1$ (resp. $\omega_j >1$) corresponds to an excitable 
(resp. oscillatory \cite{footnote}) behavior of the solitary rotator $j$. 
Throughout the paper we consider the case $\omega<1$. $D$ is the intensity of the Gaussian noises $\xi_j$ of zero mean and correlations $\left< \xi_j(t) \xi_k(t)\right> = \delta(t-t') \delta_{jk}$, and $C$ is the coupling intensity.

To characterize collective behavior we use the time-dependent global amplitude, 
$\rho(t)$, and phase, $\Psi(t)$ \cite{kmoto84,ptp:86:shinomoto}.
\begin{equation}\label{eq:rho}
\rho(t)e^{i \Psi(t)} = \frac{1}{N}\sum_{k=1}^N 
e^{i \phi_k(t)} .
\end{equation} 
The Kuramoto order parameter $\rho \equiv \langle \rho(t) \rangle$, 
where $\left< \cdot \right>$ denotes the time average, is
known to be a good measure of collective synchronization in coupled oscillators
systems, i.e. $\rho =1$ when oscillators synchronize $\phi_j(t)=\phi_k(t),\,\forall j,k$, and $\rho \rightarrow 0$ for desynchronized behavior. 
Notice, however, that the Kuramoto parameter adopts a non-zero value even when all the variables $\phi_j$, being equal to each other, are at rest. To discriminate between this static entrainment from the dynamic entrainment (time synchronization) of excitable systems when all units fire synchronously, we use the order parameter introduced by Shinomoto and Kuramoto~\cite{ptp:86:shinomoto}
\begin{equation}\label{eq:zeta}
\zeta = \left< \left| \rho(t) e^{i\Psi(t)} - 
\left< \rho(t)\, e^{i\Psi(t)} \right> 
\right|  \right>,
\end{equation}
which differs from zero only in the case of synchronous firing. 
Finally, a measure for the activity of the units, widely used in problems of stochastic transport in non-symmetric potentials is the current
\begin{equation}\label{eq:j}
J=\frac{1}{N}\sum_{k=1}^N\left< \dot\phi_k(t)\right>.
\end{equation}
A non-zero current $J$ describes a situation in which the systems are firing (not necessarily synchronized). 

We now provide an analytical theory to understand the behavior of
$\rho$, $\zeta$ and $J$ as a function of the control parameters, 
$C$, $D$ and $\sigma$. The theory proceeds in three steps. First, under 
the assumption of entrainment, we derive a dynamical equation for the 
global phase $\Psi$, depending on the value of the Kuramoto parameter $\rho$. 
Second, using the solution of that equation, we obtain expressions for 
$\zeta$ and $J$ which depend on $\rho$. Finally, we calculate 
self-consistently the value of $\rho$. 

Averaging equation (\ref{eq:phidot}) over the whole ensemble and using the definition of global amplitude and phase of 
Eq.~(\ref{eq:rho}) we have
\begin{equation}\label{eq:MF}
\frac 1 N \sum_{k=1}^N \dot{\phi_j} = \omega -  \rho(t) 
\sin \Psi(t)  +  \sqrt{\frac{ D}{N}} \xi(t).
\end{equation}
where $\xi(t)$ is a Gaussian noise of zero mean and correlations 
$\langle \xi(t) \xi(t')\rangle = \delta(t-t')$.
Taking the time-derivative of Eq.~(\ref{eq:rho}) and introducing $\delta_j(t)= \phi_j(t)- \Psi(t)$, we obtain:
\begin{equation}
\dot{\rho}(t)+ i \, \rho(t) \dot{\Psi}(t) = \frac{i}{N} \sum_{k=1}^N \dot{\phi}_k 
e^{i \delta_k (t)}.
\end{equation}
We consider now that the rotators are synchronized in the sense that $\delta_j(t)\ll 1$ 
and substitute the expansion $e^{i \delta_k}=1+ i \delta_k + 
\mathcal{O}(\delta_k^2)$ in the previous expression. Equating real and imaginary parts, we obtain
\begin{equation}
\label{preveq}
\rho(t) \dot{\Psi}(t) =  \frac 1 N \sum_{k=1}^N \dot{\phi}_k + 
\mathcal{O}(\delta_k^2).
\end{equation}
The definition of $\delta_i$ leads to $\rho(t)=N^{-1}\sum_k e^{i \delta_k}$. Hence $\dot\rho(t)=\mathcal{O}(\delta_k^2)$ and, consistently with the order of the approximation, we can replace in equation (\ref{preveq}) the time dependent $\rho(t)$ by the constant value $\rho$. Therefore, Eq.~(\ref{eq:MF}) can be approximated by $\rho \dot{\Psi}(t) = \omega - \rho \sin \Psi(t) + \sqrt{{ D}/{N}} \xi(t),$
which in the limit $N \rightarrow \infty$, reduces to
\begin{equation}\label{eq:PsiDot}
\dot{\Psi}(t) = \frac{\omega}{ \rho }-\sin \Psi(t).
\end{equation}
It is remarkable that the global phase obeys the same dynamics than the
individual units but with a natural frequency scaled with $\rho$, the 
Kuramoto parameter measuring the entrainment degree. 
Therefore, a decrease in the entrainment lowers the global excitability 
threshold from $\omega=1$ to $\omega=\rho$
and the system can start firing synchronously. The effect can be 
understood as a broadening of the distribution of the phases $\phi$, 
so that a fraction of the rotators crosses over the threshold and, if the coupling
is large enough, they pull a macroscopic fraction of the oscillators.
Thus degradation of the
entrainment has the paradoxical effect of increasing the coherent firing. 
It is essential to realize that Eq. (\ref{eq:PsiDot}) depends only on the value of $\rho$ and not in the specific way the degradation of $\rho$ is achieved, so that similar 
effects can be achieved either increasing the noise, either decreasing the coupling, or increasing the diversity in the natural frequencies; a significantly insightful result not previously understood nor discussed.

Let us now express $\zeta$ and $J$ as a function of the Kuramoto order 
parameter $\rho$. In the case $\rho < \omega$, the solution of 
(\ref{eq:PsiDot}) is given by \cite{initialcondition}
\begin{equation}
\omega-\rho \sin \Psi(t) =\frac{\omega^2-\rho^2}{\omega-\rho \cos \Omega t},
\label{eq:sin_psi}
\end{equation}
where $\Omega=\sqrt{(\omega/\rho)^2-1}$ is the frequency of the global phase 
oscillations. The current is obtained from Eq. 
(\ref{eq:MF}), $J = \omega - \left< \rho \, \sin (\Psi) \right>$. 
Time averages are computed over a period $T=2\pi/\Omega$ using 
Eq. (\ref{eq:sin_psi}) 
\begin{equation}
J = {\omega^2 - \rho^2 \over T} \int_0^T  
{dt \over \omega - \rho \cos \Omega t} = \sqrt{\omega^2-\rho^2}.
\label{eq:J_prediction}
\end{equation}
For $\rho>\omega$, $J=0$.

Approximating $\rho(t)$ by a constant value, the Shinomoto-Kuramoto 
parameter $\zeta\cong \rho \left< \left| e^{{\mathrm{i}} \Psi(t)} - 
\left< e^{{\mathrm{i}} \Psi(t)} \right>  \right|  \right>$ can be 
computed performing again the time averages over a period $T$ using 
Eq. (\ref{eq:sin_psi}):
\begin{equation}
\zeta = {2 \over \pi} 
\sqrt{2(\omega-\sqrt{\omega^2-\rho^2})(\omega+\rho)} K 
\left({2 \rho \over \rho-\omega} \right),
\label{eq:zeta_prediction}
\end{equation}
where $K(m)$ is the complete elliptic integral of the first kind \cite{abramowitz}. If $\rho>\omega$ we get $\zeta=0$.

\begin{figure*}
\begin{center}
\includegraphics[width=8.2cm,angle=-90]{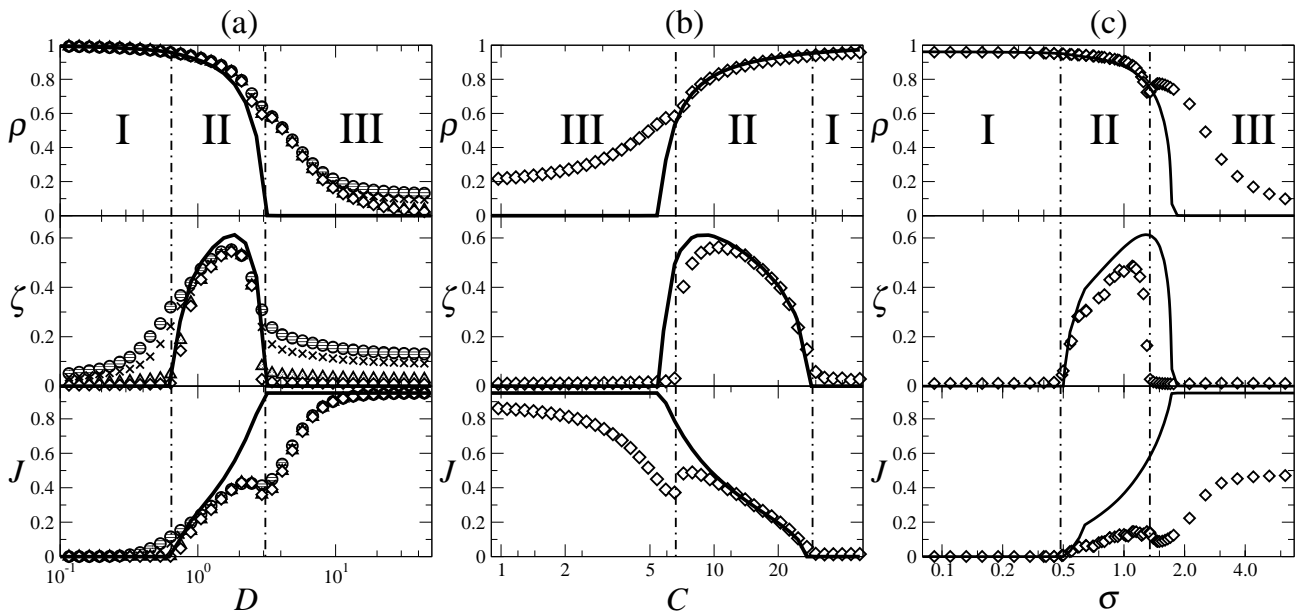}
\caption{\label{fig:ordprm} Symbols represent $\rho$, $\zeta$ and $J$ as obtained numerically from Eqs.~(\ref{eq:phidot}). Solid lines are the theoretical results. Panel (a) shows the variation with respect to the noise intensity $D$ in absence of diversity,
$\sigma=0$, for frequency $\omega=0.95$, coupling $C=4$, and different system sizes:
$N=50$ ($\circ$), $N=10^2$ ($\times)$, $N=10^3$ ($\triangle$), $N=10^4$ ($\diamond$). Panel (b) displays the same results as a function of $C$ for $D=1.0$. Panel (c) shows the variation with respect to diversity $\sigma$ for $D=0.3$, $C=4$ and $g(\omega_j$ being a uniform distribution. In all cases there are three regimes: (I) no firing, (II) synchronized firing and (III) unsynchronized firing.}
\end{center}
\end{figure*}

As a final step, we derive a equation for $\rho$ using a self-consistent, 
Weiss-like, mean field approximation, which assumes constant values 
for the global magnitudes and then averages over their probability 
distribution \cite{bpt94,kmoto84}. For our particular case, we start 
by rewriting Eq.~(\ref{eq:phidot}) as
\begin{equation}
\dot \phi_i(t) = -{dV(\phi_i;\Psi,\rho,\omega_i)\over d\phi_i} + \sqrt{D}\, \xi_i(t),
\end{equation}
where we have defined the {\sl potential}
\begin{equation}
V(\phi;\Psi,\rho,\omega)= -\omega \phi - \cos (\phi) - C\rho \cos (\Psi - \phi).
\nonumber
\end{equation}
Note that the coupling appears only through the global parameters $\rho$ 
and $\Psi$. For fixed $\rho$ and $\Psi$, the stationary probability 
distribution function reads \cite{gardiner}
\begin{equation}\label{pdf}
P_{\rm st}(\phi;\Psi,\rho,\omega)=Z^{-1} e^{-2V(\phi)/D} \int_0^{2\pi} d\phi'\,
e^{2V(\phi'+\phi)/D} ,
\end{equation}
where $Z$ is a normalizing constant. From its definition, we have 
$\rho= (1/N) \sum_{k=1}^N \left<\cos (\phi_k - \Psi) \right>$, and we obtain
\begin{equation}
\begin{array}{lcr}
\rho  = &&\label{eq:self_consistent}
\\
\int
d\omega g(\omega)\int_0^{2\pi} d\Psi P(\Psi;\rho)\int_0^{2\pi} d\phi 
P_{\rm st}(\phi;\Psi,\rho,\omega) \cos(\phi-\Psi) &&
\end{array}
\end{equation}
where we have performed a triple average: with respect to the distribution (\ref{pdf}),
with respect to the distribution $g(\omega)$ of natural frequencies and with respect to the
distribution $P(\Psi;\rho)$ of the global phase which
is inversely proportional to the instantaneous velocity given by the dynamics 
(\ref{eq:PsiDot}), namely   $P(\Psi;\rho)=(1/2\pi)\sqrt{\omega^2-\rho^2}/(\omega
- \rho \sin \Psi)$ for $\rho<\omega$ and  $P(\Psi;\rho)=\delta(\Psi-
\arcsin(\omega / \rho))$ otherwise. The self-consistent equation
(\ref{eq:self_consistent}) for $\rho$ needs to be solved numerically.

In the following, we discuss the theoretical results and compare  them with the numerical
results obtained from a numerical integration of Eqs.~(\ref{eq:phidot}).  Fig.
\ref{fig:ordprm}a shows $\rho$, $\zeta$ and $J$ as function of the noise intensity $D$  in
absence of diversity. The solid lines correspond to the theoretical results while symbols 
show the numerical results for different system sizes. In this figure, we can observe the
three aforementioned behaviors: For small noise intensity (regime I) each rotator fluctuates
around its fixed point. Although for un-coupled rotators noise would eventually excite some
spontaneous random firings, the coupling of a large number of units suppresses these
individual firings. The Kuramoto parameter $\rho$ close to $1$ and the deviations from unity
are due to the small dispersion induced by noise.  Region I is, in fact, characterized by
$\rho>\omega$ for which our theory predicts that the Shinomoto-Kuramoto parameter $\zeta$ 
and the current $J$ vanish which physically reflects the nonexistence of collective movement.
In this region, the  numerical results for $\rho$, $\zeta$ and $J$ are in excellent
agreement  with the theoretical predictions.

Our theory predicts that a transition to a dynamical state
characterized by synchronized firing behavior (regime II) takes place when 
$\rho=\omega$, in very good agreement with the numerical results.
This transition is clearly signaled by non-vanishing values of
$\zeta$ and $J$. The prediction of $\rho$ is good for a large part of 
region II (up to values of $\rho=0.7$). Later it underestimates its
value.

For very large noise intensity, the rotators desynchronize while keeping a 
non-zero current value (regime III). Hence, the synchronized activity, as 
measured by $\zeta$ goes though a maximum as noise amplitude increases. 
Our theory predicts that the transition between regions II and III occurs 
for $\rho=0$ where $\zeta=0$ and the current takes the maximum possible 
value $J=\omega$.
Surprisingly, since the small values of $\rho$ in this transition are beyond the
assumptions of the theory, the location of the second transition is also well
predicted. Moreover, the whole shape of the Shinomoto-Kuramoto parameter $\zeta$
is well reproduced over the whole range. The maximum of Eq. 
(\ref{eq:zeta_prediction}) occurs for $\rho\approx 0.821 \omega$, which is well 
confirmed by the numerical results.  The theoretically predicted current $J$
fits the numerical values in the same range than $\rho$. Note, however, the
numerical simulations show a local maximum for the current $J$ which indicates a
local increasing in the total transport due to the coherent dynamics in the
regime II. This local maximum is not present in the theoretical approximation. 

Some of these states were already described by Kuramoto and 
Shinomoto~\cite{ptp:86:shinomoto}. By looking at the probability distribution 
of $\phi_i$, these authors identify two regions in parameter space: the 
{\sl time-periodic regime} (P) and the {\sl stationary regime} (S). Region P 
corresponds to our regime II where the order parameter $\zeta$ is different 
from zero and there is collective motion of the oscillators. Our findings allow 
us to split region S of these authors in our  distinct regions I and III: while 
region I is a fluctuating regime around the steady state, region III has a high 
activity as characterized by a non-zero current $J$. 

These results indicate that noise acts in two antagonistic ways: while a given
noise intensity can excite the sub-threshold units, forcing a synchronized
firing, large amplitude noise deteriorates the synchronization properties of
the ensemble. This scenario resembles the so called noise induced phase 
transitions~\cite{bpt94} in which a transition to an ordered 
ferromagnetic-like state is induced by increasing the noise intensity; the 
order is destroyed again for large enough noise.
Here, the transition is towards an organized collective motion of the active 
rotators.

The reverse scenario can be observed varying the coupling strength $C$, see
Fig. \ref{fig:ordprm}b. The Kuramoto parameter $\rho$ increases 
with $C$ (notice, however, the existence of a small bump in the numerical results),
indicating that the degree of synchronization increases with coupling, as
expected. A large coupling suppresses noise-induced firings, and the system is
macroscopically at rest, regime I, as indicated by the vanishing of $\zeta$ and 
$J$. For weak coupling the noise induces desynchronized individual firings 
(regime III) characterized by a macroscopic current $J$ and again a zero value 
for $\zeta$. For intermediate values of the coupling (regime II) the interplay 
of noise and coupling leads to the largest degree of synchronized firing with 
a large value for $\zeta$.

Finally Fig.\ref{fig:ordprm}c shows $\rho$, $\zeta$ and $J$ as a function of 
the diversity $\sigma$. It is clear in the figure the existence of the same three 
regimes that were obtained  by varying the noise intensity or the coupling. 
Altogether Fig. \ref{fig:ordprm} clearly illustrates the fact 
that similar effects can be achieved increasing the noise, decreasing the 
coupling or increasing the diversity in the natural frequencies as theoretically
predicted.

In summary we have developed a theory for the emergence of coherent firing in  subthreshold
excitable systems. It arises as a consequence of a phase transition induced by noise or
diversity.  The underground mechanism is the degradation of entrainment originated by
disorder. This leads to the somehow paradoxical result of  establishing a lower effective
threshold for collective firing.  It does not matter the specific source of disorder
responsible for  entrainment degradation, either noise or diversity when nonlinearly 
combined with coupling lead to similar results. Coherent firing is thus  achieved by a global
saddle-node bifurcation, triggered by the degree of  entrainment of the ensemble, $\rho$.
This mechanism is not restricted to the model we considered,  it will exist in any physical,
chemical or biological system with the  necessary generic ingredients.

We acknowledge financial support by the Ministerio de Educaci\'on y Ciencia (Spain), FEDER projects FIS2004-5073, FIS2004-953, BFM2001-0341 and the EU NoE BioSim (LSHB-CT-2004-005137).

\bibliography{refs}

\begin{thebibliography}{21}
\bibitem[{foo()}]{footnote}
\bibinfo{note}{In this case, note that actual frequency is
  $\sqrt{\omega^2-1}$.}
\bibitem[{ini()}]{initialcondition}
\bibinfo{note}{This form of the solution assumes a particular initial condition
  $\Psi(t=0)=-\pi/2$, which is otherwise irrelevant when taking the time
  average.}
\end{thebibliography}

\end{document}